# Quantitative characterization of highly efficient correlated photon-pair source using biexciton resonance

**Yasuo Yamamoto, Goro Oohata[*], and Kohji Mizoguchi**

*Department of Physical Science, Osaka Prefecture University, 1-1 Gakuen-cho, Naka-ku, Sakai-shi, Osaka 599-8531, Japan*
[*]*oohata@p.s.osakafu-u.ac.jp*

**Abstract:** A high efficiency method for the generation of correlated photon pairs accompanied by a reliable means to characterize the efficiency of that process is needed in the study of correlated photon pairs and entangled states involving more than three photons, which have important potential applications in quantum information and quantum communication. In this study, we report the first characterization of the efficiency of generation of correlated photon pairs emitted from a CuCl single crystal using the biexciton-resonance hyper-parametric scattering (RHPS) method which is the highly efficient method of generation of correlated photon pairs. In order to characterize the generation efficiency and signal-to-noise ratio of correlated photon pairs using this method, we investigated the excitation density dependence on the photon counting rate and coincidence counting rate under resonant excitation. The excitation density dependence shows that the power characteristics of the photon counting rates changes from linear to quadratic dependence of the excitation density. This behavior represents a superposition of contributions from correlated photon pairs and non-correlated photons. Photon counting signals in this study were recorded as time-tag data, which provide the photon counting rate and coincidence counting rate simultaneously, and improve the coincidence counting rate by the factor of 200 compared to that obtained in previous work [K. Edamatsu *et al*. Nature **431**, 167-170 (2004)]. The analysis of the excitation density dependence shows that one photon-pair is produced by a pump pulse with 2 x $10^6$ photons. Moreover, the generation efficiency of this method is improved by a factor of $10^7$ compared to that of several methods based on the $\chi^{(3)}$ parametric process.

**1. Introduction**

Quantum entangled photon pairs are essential to realize various quantum-information processing protocols such as quantum teleportation [1] and quantum cryptography [2]. Entanglement shared by more than three particles, which is referred to as "multipartite entanglement," is a significant resource for quantum simulation [3-5], quantum lithography [6], and quantum computation [7-9]. In order to generate multipartite entanglement among a large number of photon pairs, a high pump-power, highly efficient detection systems, and highly efficient sources of entangled photon pairs are required [10, 11].

The generation of entangled photon pairs from a CuCl single crystal via biexciton-resonance hyper-parametric scattering (RHPS) has previously been reported [12, 13]. The RHPS method applied to a CuCl crystal is reported to be a highly efficient method to generate entangled photon pairs. However, there are no reports that quantitatively characterize the efficiency of this method. In order to apply RHPS method using CuCl to the generation source for multipartite entanglement, it is necessary to first clarify the efficiency of the method in producing entangled photon pairs.

The RHPS method is related to a third-order nonlinear parametric process. In the RHPS method using CuCl, the third-order nonlinear susceptibility $\chi^{(3)}$ is significantly enhanced by utilizing the resonance of two-photon absorption to the biexciton state of CuCl. However, signals from the non-correlated photons, which originate from the Rayleigh scattered light and the luminescence from the bound excitons, are non-negligible. These non-correlated photons are recorded as background counts under the measurements of time-correlated histograms, which prevent the estimation of the efficiency of generating correlated photon pairs. On the other hand, the number of correlated photon pairs obtained by the RHPS method is proportional to the square of the excitation density. This results in a difference in the excitation density dependence between correlated photon pairs and non-correlated photons. Thus, the excitation density dependence of photon counting rate are expressed as a superposition of contributions from correlated photon pairs and non-correlated photons. Therefore, the investigation of the excitation density dependence enables us to estimate the generation efficiency of correlated photon pairs.

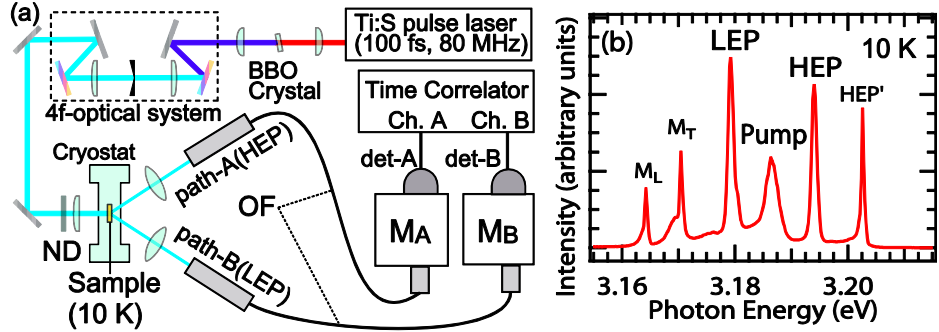

Fig. 1. (a) Schematic of the experimental setup for the resonant hyper-parametric scattering (RHPS) method. ND: neutral density filter, OF: optical multimode fiber, $M_A$ and $M_B$: monochromators, det-A and det-B: photon counting detectors (photo-multiplier tube). (b) RHPS spectrum for a CuCl single crystal. The central peak indicates the scattered light of the pump pulse (3.186 eV). The two side peaks around the pump beam are the RHPS signals of a high-energy polariton (HEP; 3.194 eV) and low-energy polariton (LEP; 3.179 eV). The peaks $M_T$ and HEP' are other RHPS signals that propagate in counter directions. The peak $M_L$ is the emission from the biexciton leaving the longuitudinal excitons.

In this paper, we report the first quantitative characterization of the generation efficiency of correlated photon pairs using the RHPS method with CuCl. The photon counting signals are obtained by a method that enables us to acquire time-tag data (TTD), which represents a simultaneous measurement of the photon counting rate and the coincidence counting rate. The efficiency of generating correlated photon pairs was characterized by investigating the excitation density dependence of the photon counting rate and the coincidence counting rate from TTD. The obtained efficiency of correlated photon pairs was at least $10^7$ times greater than those of several other methods relying on the $\chi^{(3)}$ parametric process.

## 2. Experiment

A CuCl single crystal sample with a size of approximately 5 x 5 x 0.1 $mm^3$ grown from vapor phase was used for the experiment. The temperature of the sample was maintained at 10 K in a cryostat. A schematic of the experimental setup for the RHPS method is shown in Fig. 1(a). The pump pulses were the second harmonic light of a mode-locked Ti:Sapphire pulse laser at a repetition rate of 80 MHz. For excitation at the two-photon resonance of the biexciton state in CuCl, the center energy and spectral width of the pump pulses are selected to be $\hbar\omega$ = 3.186 eV and $\Delta\hbar\omega$ = 1.6 meV, respectively, by using a 4f optical system composed of two lenses, two gratings, and a slit. The duration of the pump pulses formed through the 4f optical system was approximately 1 ps. The pump pulses passing through a neutral density filter were focused on the sample, where the spot size of the pump pulses on the sample was approximately 100 μm. The photons emitted from the CuCl single crystal were led into optical multi-mode fibers marked as path-A and path-B in conjunction with two monochromators. The angle between the direction of transmitted pump pulses and the direction of detected photons from RHPS was approximately 45°. The spectrum observed using a charge-coupled device (CCD) camera attached to a monochromator is shown in Fig. 1(b). The two side peaks around that of the pump pulse at 3.186 eV indicate the polariton modes generated through RHPS. These two peaks observed at 3.194 and 3.179 eV correspond to a high energy polariton (HEP) and low energy polariton (LEP), respectively. Moreover, the two peaks indicate the signals due to the correlated photon pairs by taking into account the phase matching condition. The photons emitted from the HEP and LEP are led to two monochromators labeled $M_A$ and $M_B$ that are connected to photo-multiplier tubes (PMTs) marked as det-A and det-B, respectively. The quantum efficiency of PMTs were approximately 30 % around 3 eV. The photon counting signals were detected by each PMT in conjunction with a time correlator. We utilized the acquisition method of TTD for time correlation. The TTD method records all the arrival times of detected photons. It has a large advantage over a time interval analyzer (TIA), because the photon counting rates and time-correlated counting rates can be simultaneously obtained using the TTD method. The temporal resolution and dead time of the TTD method used in this study are approximately 165 ps and less than 10 ns, respectively. The upper limit of the time correlation counting rate of the TTD method is higher than that of a usual TIA.

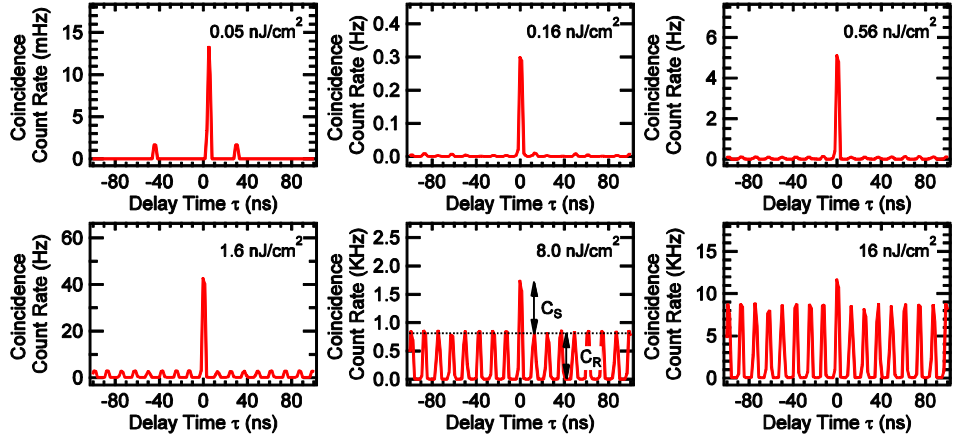

Fig. 2. Time-correlation histograms of the observed photon pairs at various excitation densities. The dashed line indicates the mean count rate of $\tau \neq 0$ and arrows indicate the value of $C_S$ and $C_R$.

### 3. Results and Discussion

The time-correlation histograms obtained at various excitation densities are shown in Fig. 2. The coincidence signals at $\tau = 0$ are clearly observed at all excitation densities. Here, the difference between the coincidence signals at $\tau = 0$ and $\tau \neq 0$, marked with $C_S$, indicates the coincidence counting rate of correlated photon pairs. The average value of the coincidence signals at $\tau \neq 0$, marked with $C_R$, indicates the coincidence counting rate of accidental photons due to background photons and photon pairs. Here, the background photons are the non-correlated photons that originate from the luminescence of bound excitons and photons due to Rayleigh scattered light at the samples, and the number of background photon shows a linear dependence on excitation density. The time-correlation histograms at various excitation densities clearly illustrate that as the excitation density increase, the values of $C_S$ and $C_R$ increase, while the true-coincidence to accidental-coincidence ratio ($C_S/C_R \equiv CAR$) decreases. The obtained variation of $CAR$ with an increase in the excitation density is consistent with the theoretical variation of $CAR$ which shows that the value of $CAR$ decreases with increasing numbers of correlated photon pairs [14]. At the excitation density of 1.6 nJ/cm$^2$, the value of $C_S$ obtained using the TTD method is approximately 200 times greater than that obtained using a TIA and a pulse laser with a repetition rate of 80 MHz [12]. Moreover, the maximum value of $CAR$ is approximately 80, which is four times that obtained using a TIA and a pulse laser with a higher repetition rate of 1 GHz [13]. These results indicate that the simultaneous measurement of photon counting rates and coincidence counting rates using the TTD method is well capable of obtaining the coincidence signals in the wide range of 0.05 – 16 nJ/cm$^2$.

We estimated the generation efficiency of correlated photon pairs from the photon counting rates and coincidence counting rates obtained by the RHPS method using CuCl. Since the background photons emitted from the CuCl single crystal is non-negligible, it prevent from estimating the generation efficiency of correlated photon pairs from only an observation of photon counting rate and coincidence counting rate. Then, we estimated the generation efficiency of correlated photon pairs from the excitation density dependence of a number of generated photons. The number of correlated photon pairs $g_{\text{pair}}$ is proportional to the square of the excitation density $I$ because the RHPS process is related to the third-order non-linear parametric process. However, the number of background photons $g_{bg}$ is proportional to the excitation density, as mentioned above. $g_{pair}$ and $g_{bg}$ are expressed as

$$g_{pair} = \alpha I^2,$$
$$g_{bg;A} = \beta_A I, \quad (1)$$
$$g_{bg;B} = \beta_B I,$$

where $\alpha$ is the generation efficiency of the correlated photon pairs and $\beta_A$ and $\beta_B$ are that of the background photons emitted in the direction of path-A and path-B, respectively. When the total detection efficiency for the directions of path-A and path-B are denoted by $\eta_{S;A}$ and $\eta_{S;B}$, the photon counting rate obtained by each detector of det-A and det-B are represented by

$$C_A = \eta_{S;A} \left( \alpha I^2 + \beta_A I \right),$$
$$C_B = \eta_{S;B} \left( \alpha I^2 + \beta_B I \right). \quad (2)$$

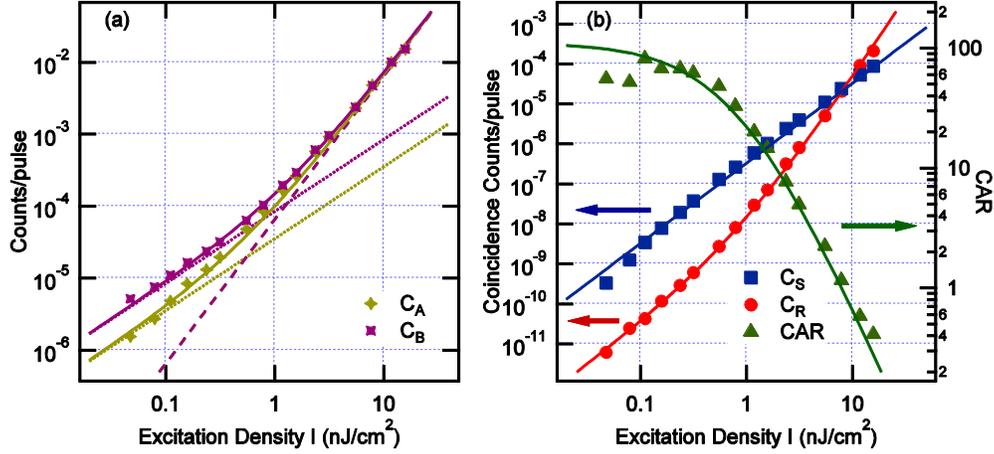

Fig. 3. (a) Excitation density dependence of photon counting rates $C_A$ and $C_B$. The solid lines show the fitted results obtained using Eq. (2). The dotted and dashed lines show the linear and quadratic functions, respectively. (b) Excitation dependence of coincidence counting rates $C_S$, $C_R$, and $CAR$. The blue solid line is the fitted result. The red and green solid lines are the results obtained using Eq. (5) and (6).

This equation indicates that the photon counting rates obtained by the respective detectors are expressed by the superposition of the excitation density dependence of the correlated photon pairs and that of the background photons, which enables us to quantitatively estimate the generation efficiency of the correlated photon pairs. The photon counting rates detected in the directions of path-A and path-B are plotted as a function of excitation density in Fig. 3(a). As the excitation density is increased, the power characteristic of the photon counting rates gradually changes from a linear to quadratic dependence of the excitation density. The generation efficiencies are estimated by fitting Eq. (2) to the excitation density dependence. For this fitting, the total detection efficiency for each detection of path-A and path-B was set to $\eta_S = \eta_{S;A} = \eta_{S;B} = 2.5\ \%$, which was estimated from losses of optical components and quantum efficiency of PMTs. The fitted results are in good agreement with the experimental results. From the fitting procedure, the generation efficiency of the correlated photon pairs was estimated as $\alpha = 2.6 \times 10^{-3}$ pairs/pulse/(nJ/cm$^2$)$^2$, which demonstrates that one correlated photon pair is produced by a pump pulse that contains $2 \times 10^6$ photons. The high generation efficiency is caused by the excitation at the resonance energy of the two-photon absorption to the biexciton state of CuCl. We will later discuss a comparison of the generation efficiencies by the RHPS method and several generation methods that use the $\chi^{(3)}$ parametric process.

Moreover, the generation efficiencies of the background photons were $\beta_A = 1.4 \times 10^{-3}$ photons/pulse/(nJ/cm$^2$), and $\beta_B = 3.4 \times 10^{-3}$ photons/pulse/(nJ/cm$^2$), respectively. The generation efficiency of the background photons in the direction of path-B is slightly larger than that for path-A. This difference is caused by the luminescence from the bound excitons at 3.180 eV.

Next, we discuss the excitation density dependence of the coincidence counting rate. The time-correlation function of the photon counting rates for det-A and det-B is represented as a function of the delay time τ between photons as

$$G^{(2)}(\tau) = \langle C_A(t) C_B(t+\tau) \rangle. \tag{3}$$

Here, $C_A(t)$ and $C_B(t)$ indicate the photon counting rates as functions of time $t$ by the detectors of det-A and det-B, respectively. Using the time-correlation function and Eq. (2), the coincidence counting rates of the correlated photon pairs $C_S$, the accidental photons $C_R$, and the value of $CAR$ are given as a function of excitation density $I$ by [15]

$$C_S(I) = G^{(2)}(0) - G^{(2)}(\infty) = \eta_X \eta_S^2 \alpha I^2, \tag{4}$$

$$C_R(I) = G^{(2)}(\infty) = \eta_S^2 \left( \alpha I^2 + \beta_A I \right)\left( \alpha I^2 + \beta_B I \right), \tag{5}$$

$$CAR(I) = \frac{C_S(I)}{C_R(I)} = \frac{\eta_X \alpha I^2}{\left( \alpha I^2 + \beta_A I \right)\left( \alpha I^2 + \beta_B I \right)}, \tag{6}$$

where $\eta_X$ is the simultaneous detection efficiency of the correlated photon pairs for both detectors, det-A and det-B. Figure 3(b) shows the excitation density dependence of coincidence counting rates for $C_S$, $C_R$, and $CAR$. The value of $C_S$ is proportional to the square of the excitation density, while the value of $C_R$ changes from the quadratic dependence to fourth power dependence with increasing excitation density. The excitation density dependence of $C_S$ and $C_R$ is consistent with that represented by Eqs. (4) and (5). The value of $\eta_X$ was estimated to be 20.3 % by fitting Eq. (4) to the excitation density dependence of $C_S$, where the parameters $\eta_S$ and $\alpha$ were set to the values estimated above. Moreover, the excitation density dependences of $C_R$ and $CAR$ were calculated by using the values of $\eta_S$, $\alpha$, $\beta_A$ and $\beta_B$ obtained above. The calculated excitation density dependences of $C_R$ and $CAR$ are in good agreement with the experimental results, which confirms that the obtained efficiencies are valid. Here, as the excitation density decreases, the value of $CAR$ increases and converges to a maximum value at the low excitation density limit. The reason of this convergence is that the both of excitation density dependence of $C_S$ and that of $C_R$ show a quadratic dependence in the low excitation density limit. Using the low excitation density limit in Eq. (6), the maximum value of $CAR$ is expressed as

$$CAR_{max} = \frac{\eta_X \alpha}{\beta_A \beta_B}. \tag{7}$$

The value of $CAR_{max}$ was estimated as approximately 100 from the parameters obtained above. Equation (7) shows that the value of $CAR_{max}$ is given by the ratio of the generation efficiency of the correlated photon pairs to that of the background photons, which is applicable only to the generation method of correlated photon pairs using the $\chi^{(3)}$ parametric process. Then, to directly compare the generation efficiency of the RHPS method with that of several generation methods using the $\chi^{(3)}$ parametric process, we define $CAR'_{max}$ as a quantitative indicator presented by

$$CAR'_{max} = \frac{\alpha}{\beta_A \beta_B}. \tag{8}$$

In the present RHPS method, the value of $CAR'_{max}$ was obtained as $5.5 \times 10^2$, and was compared with that obtained by an inelastic four-photon scattering (FPS) method and a spontaneous four wave mixing (SFWM) method which are well-known generation methods of correlated photon pairs using the $\chi^{(3)}$ parametric process. The values of $CAR'_{max}$ obtained by the FPS method using a dispersion shifted (DS) fiber and the SFWM method using a silicon-waveguide (Si-WG) structure were estimated to be 6.5 x 10 and 9.3 x 10, respectively [16, 17]. The value of $CAR'_{max}$ obtained by the RHPS method using CuCl is several times higher than the above-reported values, which clarifies that the background photons in RHPS are relatively fewer than those of the two kind of generation methods.

Finally, we compare the generation efficiency of the correlated photon pairs by the RHPS method with that obtained by several other generation methods that use the $\chi^{(3)}$ parametric process. The generation efficiencies of correlated photon pairs by the FPS method using a DS fiber and the SFWM method using a Si-WG structure have been estimated to 1.6 x $10^{-10}$ and 2.2 x $10^{-17}$ pairs/pulse/(nJ/cm$^2$)$^2$, respectively [16, 17]. The generation efficiency of correlated photon pairs for the RHPS method using CuCl is over $10^7$ times larger than that of the FPS and SFWM methods. From this high generation efficiency, we expect that the bright correlated photon pairs are generated even by using a CW laser. Accordingly, by using a CW laser with an excitation power of 1 mW and a spot size of 100 μm, the coincidence counting rate of correlated photon pairs is estimated as 2.9 x $10^2$ Hz with a total detection efficiency of 2.5 %. This value is large enough for applications in quantum information and quantum communication that use CW lasers with a few milliwatts. Moreover, the high generation efficiency obtained in this study demonstrates that RHPS method using CuCl is a highly efficient method to generate correlated photon pairs and to investigate the multipartite entanglement consisting of a large number of photon pairs.

## 4. Conclusion

We investigated the generation efficiency of correlated photon pairs emitted from a CuCl single crystal using the RHPS method by acquisition of TTD which simultaneously provides photon counting rates and coincidence counting rates. The TTD method improves the detection efficiencies of the coincidence counting rate and the true-coincidence to accidental-coincidence ratio, which were 200 and 4 times greater than those obtained using a TIA, respectively. The generation efficiency of correlated photon pairs was evaluated by investigating the excitation density dependence of the photon counting rate and coincidence counting rate due to correlated photon pairs and the background photons. The evaluated generation efficiency of correlated photon pairs obtained by the RHPS method using CuCl is $10^7$ times larger than that obtained by several generation methods that utilize the $\chi^{(3)}$ parametric process. This result demonstrates that the RHPS method using CuCl is an extremely efficient generation method of correlated photon pairs, and is expected to be applied to the source of multipartite entangled photons and the excitation light source with non-classical properties.

## 5. Acknowledgement

This work was supported by Research Foundation for Opto-Science and Technology, Support Center for Advanced Telecommunications Technology Research, and JSPS KAKENHI Grants No. 24654092 and No. 26610088. We thank Dr. S. S. Garmon for useful comments.